\documentclass[aps,prb,twocolumn,floatfix,amsmath,amssymb,showpacs,
               superscriptaddress,10pt]{revtex4-1}
\usepackage[final]{graphicx}
\usepackage[caption=false]{subfig}
\usepackage{placeins} 
\usepackage{float}
\usepackage{color}
\usepackage{dcolumn}
\usepackage{xcolor}
\usepackage{url}
\graphicspath{{Figures/}}
\usepackage{mathtools}
\usepackage{bm}
\usepackage{epsfig,psfrag,amsmath,amssymb}
\input{epsf}
\usepackage{hyperref}
\usepackage[percent]{overpic}

\hypersetup{
    colorlinks=true,
    linkcolor=blue,
    citecolor=blue,
    filecolor=magenta,      
    urlcolor=blue,
}
\usepackage{array}
\usepackage{tabu}
\newcolumntype{C}[1]{>{\centering\arraybackslash$}p{#1}<{$}}

\setcitestyle{square}

\bibpunct{[}{]}{,}{n}{}{}
 
\begin{document}
\title{Range of biquadratic and triquadratic Heisenberg effective couplings deduced from
multiorbital Hubbard models}

\author{Rahul Soni}
\affiliation{Department of Physics and Astronomy, The University of 
Tennessee, Knoxville, Tennessee 37996, USA}
\affiliation{Materials Science and Technology Division, Oak Ridge National 
Laboratory, Oak Ridge, Tennessee 37831, USA}
\author{Nitin Kaushal}
\affiliation{Materials Science and Technology Division, Oak Ridge National 
Laboratory, Oak Ridge, Tennessee 37831, USA}
\author{Cengiz \c{S}en}
\affiliation{Department of Physics, Lamar University, Beaumont, Texas, 77710, USA}
\author{Fernando A. Reboredo}
\affiliation{Materials Science and Technology Division, Oak Ridge National 
Laboratory, Oak Ridge, Tennessee 37831, USA}
\author{Adriana Moreo}
\affiliation{Department of Physics and Astronomy, The University of 
Tennessee, Knoxville, Tennessee 37996, USA}
\affiliation{Materials Science and Technology Division, Oak Ridge National 
Laboratory, Oak Ridge, Tennessee 37831, USA}
\author{Elbio Dagotto}
\affiliation{Department of Physics and Astronomy, The University of 
Tennessee, Knoxville, Tennessee 37996, USA}
\affiliation{Materials Science and Technology Division, Oak Ridge National 
Laboratory, Oak Ridge, Tennessee 37831, USA}
\date{\today}

\begin{abstract}
We studied a multi-orbital Hubbard model at half-filling for two and three orbitals per site on a two-site cluster via full exact diagonalization, in a wide range for the onsite repulsion $U$, from weak to strong coupling, and multiple ratios of the Hund coupling $J_H$ to $U$. The hopping matrix elements among the orbitals were also varied extensively.  At intermediate and large $U$, we mapped the results into a Heisenberg model. For two orbitals per site, the mapping is into a $S=1$ Heisenberg model where by symmetry both nearest-neighbor $(\mathbf{S}_{i}\cdot\mathbf{S}_{j})$ and $(\mathbf{S}_{i}\cdot\mathbf{S}_{j})^{2}$ are allowed, with respective couplings $J_1$ and $J_2$. For the case of three orbitals per site, the mappping is into a $S=3/2$ Heisenberg model with $(\mathbf{S}_{i}\cdot\mathbf{S}_{j})$, $(\mathbf{S}_{i}\cdot\mathbf{S}_{j})^{2}$, and $(\mathbf{S}_{i}\cdot\mathbf{S}_{j})^{3}$ terms, and respective couplings $J_1$, $J_2$, and $J_3$. The strength of these coupling constants in the Heisenberg models depend on the $U$, $J_H$, and hopping amplitudes of the underlying Hubbard model. Our study allows to establish bounds on how large the ratios $J_2/J_1$ and $J_3/J_1$ can be. We show that those ratios are severely limited and, as a crude guidance, we conclude that $J_2/J_1$ is less than 0.4 and $J_3/J_1$ is less than 0.2, establishing bounds on effective models for strongly correlated Hubbard systems. 
\end{abstract}
\maketitle

\section{Introduction}

The study of the one-dimensional spin-one ($S=1$) Heisenberg chain by Haldane~\cite{haldane}, with only nearest-neighbor spin-spin interactions (called here ``quadratic'' interactions), and the prediction, and subsequent confirmation, of a spin liquid gapped ground state with protected edge states, was seminal for the start of the field of topological materials. The Haldane chain has been physically realized in several materials, such as
CsNiCl$_3$~\cite{CNC}, AgVP$_2$S$_6$~\cite{AVPS}, NENP~\cite{NENP}, and Y$_2$BaNiO$_5$~\cite{YBNO}, and
recently theory predicted that doping of the fermionic two-orbital Hubbard 
version of the idealized Haldane chain may lead
to hole pairing and eventual superconductivity~\cite{njpDagotto,nirav17}. Earlier 
related work employing $t-J$ model approximations
also predicted superconductivity with doping although strongly competing with ferromagnetism~\cite{riera97}.

While the solution of the Heisenberg $S=1$ chain by Haldane was mathematically elegant, intuition was provided later
by Affleck et al.~\cite{affleck} when they solved exactly an extension of the original quadratic Hamiltonian by
adding ``biquadratic'' terms. In this exactly solvable point, the magnitude of the
ratio biquadratic to quadratic couplings is $\beta = 1/3$~\cite{affleck}. At this special point, the model has properties qualitatively similar to those of the Haldane chain, with a unique spin-gapped
ground state, exponentially decaying spin-spin correlations, and $S=1/2$ spins at the edges when open boundary conditions are used.

The primary goal of the investigations reported in this publication 
is to study whether the more realistic electronic two-orbital Hubbard model
realization of the Haldane chain recently introduced~\cite{njpDagotto} can at large and/or intermediate Hubbard $U$ and Hund $J_H$ couplings reach the biquadratic/quadratic ratio $\beta=1/3$ when fermionic vs. pure spin Hamiltonian models are compared at low energies. Specifically, here we solve exactly the two-site problem of the fermionic model and represent the lowest energy states using the generalized Heisenberg quadratic-biquadratic model in a vast region of
parameter space, including varying the elements of the hopping matrix. Our conclusion is that it is indeed possible to reach the Affleck et al. point by suitably selecting
the values of $U$ and $J_H$. On the other hand, for the Bethe-Ansatz solvable case $\beta=1$ we conclude that it cannot be reached using the fermionic system we studied. Our efforts were extended to the 
three-orbital per site Hubbard models as well, allowing us to deduce upper bounds 
for the biquadratic and triquadratic Heisenberg couplings emerging at large Hubbard $U$ and low energy. 

\section{Previous investigations using quadratic-biquadratic $S=1$ spin models}

Interest in spin Heisenberg models with spin higher than 1/2 started developing years ago in the context of finding exactly solvable Hamiltonians, in dimension one or  more, to uncover
disordered spin liquid ground states in antiferromagnets. Of particular interest were valence bond (VB) states, which could serve as toy models for the ideas of Anderson using $S=1/2$ resonant valence bonds related to high-$T_c$ superconductivity~\cite{anderson}. Affleck and collaborators 
extended the notion of VB states to spins higher than 1/2~\cite{affleck}, as explained in the introduction. For $S=1$ adding a biquadratic nearest-neighbor term with coupling $J_2$ in addition to the standard (quadratic) Heisenberg interaction with coupling $J_1$, they found that for $\beta=J_2/J_1=1/3$ (with $J_1$ and $J_2$ both positive, thus antiferromagnetic) the ground state is exactly solvable and indeed made out of
valence bonds. The same model but for the case  $|\beta|=J_2/J_1=1$ has been 
solved using the Bethe Ansatz method~\cite{tak}. At this special point the ground state is gapless with a power-law decay. This point, with $J_1>0$ and $J_2<0$, could separate the spin liquid gapped phase for 
$\beta>-1$ from a dimerized phase for $\beta<-1$. Our analysis below indicates that Affleck et al.'s 
case $\beta=1/3$ indeed can be realized with a two-orbital per site electronic model 
at intermediate Hubbard $U$, but the ratio $|\beta|=1$ is too large and will require more general electronic models. 

To summarize, the isotropic $S=1$ Heisenberg model with a biquadratic term has been previously studied. The phase diagram of the model in 1D was obtained via DMRG~\cite{jolicoeur}. These authors verified 
that for $\beta=1/3$ the ground state is indeed a VB state. In addition they obtained the following phases: 
(i) For $J_1>0$ and $J_2=0$ the system has a non-degenerate disordered ground state with 
antiferromagnetic spin correlations that decay exponentially indicating the presence of a gap in the spectrum (i.e. the Haldane state); 
(ii) at $\beta=1/3$ with $J_1$ and $J_2$ both positive, the system 
has the VB ground state with a spin gap in the spectrum, as described before (i.e. the Affleck et al. state); 
(iii) $\beta=1$ with both $J_1$ and $J_2$ positive, indicates the critical point where the Hamiltonian is
integrable with a gapless ground state~\cite{itoi}; 
(iv) for $\beta=1$ with both $J_1$ and $J_2$ negative, the model is also integrable and it has 
a gapless ground state;  
(v) for $J_1=0$ and $J_2=-1$ the system is in a dimerized state; this dimerized state is 
the ground state when $J_2<0$  and $|\beta|\ge 1$; finally, 
(vi) the ground state is FM for $J_1=0$ and for $\beta<-1$ and for $J_1<0$ and $|\beta|\le 1$. The region between $J_1>0$ and $|\beta|\le 1$ has a gapped 
ground state~\cite{millet}.
Several of these phases have been theoretically confirmed via a variety of approaches such as calculation of
static and dynamic structure factors using recursion methods~\cite{schmitt}. 

Moreover, recent efforts by one of the present coauthors and collaborators 
searched for spin liquids in two dimensions
focusing on the SU(3) point where the strength of the quadratic and biquadratic interactions are equal, and adding further interactions~\cite{qimiao1,qimiao2}.
Indeed spin liquids were unveiled for the spin-only models, a conceptually interesting result. 
But, as already explained, it is difficult to establish which electronic fundamental multiorbital model can realize these complex spin models in the limit of large $U$, with the exception of the state of Affleck et al. which 
is reachable with the two-orbital per site model studied here.
Our investigations establish limits
based on basic Hubbard models on what range of $\beta$ could be realized in practice. For larger
values of $\beta$ more complex fermionic models will be required.

In addition, it was shown that for certain values of parameters higher 
spin Heisenberg Hamiltonians in one dimension posses conformal invariance, property that allows an analytical determination of critical exponents~\cite{conf}. 
The integrable high-spin Heisenberg models are given by a Hamiltonian with a polynomial form in powers of nearest-neighbor Heisenberg interactions ranging from 1 to $2S$.
This was demonstrated via a mapping into the Wess-Zumino-Witten model at specific values of the Hamiltonian parameters~\cite{affleck2,affleck3}. Various numerical studies of 
higher spin Heisenberg Hamiltonians were performed to understand whether the higher spin anisotropic Heisenberg Hamiltonians belong to the same universality class as the $S=1/2$ 
isotropic model or, instead, the isotropic integrable higher spin ones~\cite{moreo,alcaraz,hallberg}.  

The isotropic $S=3/2$ Heisenberg model has not been as much explored. The isotropic and anisotropic cases were studied in order to determine whether they belong to the same universality class as 
the $S=1/2$ Heisenberg model, which was confirmed using Lanczos and DMRG approaches~\cite{affleck,moreo,hallberg, alcaraz}. 
%
Other authors have explored the $S=1$ biquadratic model in two dimensions using DMRG in the context of high-$T_c$ superconductors finding nematic phases~\cite{qimiao1}, although at robust $J_2/J_1$. 


Spin 1 systems are also realized in the area of two dimensional 
ruthenates~\cite{hotta2001} often using three-orbital per site Hubbard
models with four electrons in those three orbitals, leading to a net $S=1$ per site. 
Rich phase diagrams were reported. But in these ruthenates $S=1$ effective Hamiltonians are rarely employed. Spin 1
systems often appear also in the area of iron superconductors because Fe$^{2+}$, 
with $n=6$ electrons in the $3d$ shell, is the usual iron valence, either in planes or ladders. However, these iron-materials are considered to reside in the intermediate $U$ region~\cite{jacek1,jacek2} and, again, they are not often described via purely spin systems but with multi-orbital electronic models instead~\cite{Ni-superconducting}.

Our study also has limitations. For example, the addition of a Zeeman magnetic term to the biquadratic $S=1$ model was explored using DMRG~\cite{okunishi}, and 
a spin nematic phase was observed in a triangular lattice~\cite{tsune}. The addition of single-ion anisotropy to the $S=1$ spin Heisenberg model was studied using quantum Monte Carlo and series 
expansions~\cite{albuquerque}, and for the model with biquadratic term~\cite{dechiara} with density matrix renormalization group (DMRG). The addition of a next-nearest neighbor term to the $S=1$ Heisenberg model with biquadratic coupling
was explored as well with DMRG~\cite{pixley}. More recently, research on this model has focused on entanglement and topological properties~\cite{thomale,golinelli}. Because our study relates to a single bond, we cannot distinguish
between square and triangular lattices. Including more than a single bond, terms such as $({{{\bf S}_i}\cdot{{\bf S}_j}})({{{\bf S}_i}\cdot{{\bf S}_k}})$ with sites $(i,j,k)$ belonging to the same plaquette, will also appear in the large $U$ expansion.
Note that the models described in this paragraph have  either a
Zeeman term, single-ion anisotropy, or next-nearest neighbor interactions. Thus, it is too early to make statements
on whether these models can or cannot be realized with fermionic two-orbital Hubbard models. 
As a consequence, our study should be considered qualitative, but it still provides a crude but valuable estimation 
of how large some extra terms beyond the canonical quadratic Heisenberg interactions can be.

\section{Model and Method}

\subsection{Multi-Orbital Hubbard Model}
\noindent
For the exact-diagonalization calculations, we work with the multi-orbital Hubbard model mentioned in \cite{njpDagotto,nirav17} and described as follows:
\begin{eqnarray}
H_{H} &=& -\sum_{\substack{\langle i,\gamma; j,\gamma'\rangle ; \sigma}}t_{\gamma\gamma'} \left(c^{\dagger}_{i,\gamma,\sigma}c_{\vphantom{i^i}j,\gamma',\sigma} + h.c\right)  \nonumber \\
&+& U \sum_{i,\gamma} n_{i,\gamma,\uparrow}n_{i,\gamma,\downarrow} + \left(U'-\frac{J_{H}}{2}\right)\sum_{\substack{i \\ \gamma < \gamma'}}n_{i,\gamma}n_{i,\gamma'}  \nonumber \\
&-& 2J_{H}\sum_{\substack{i \\ \gamma < \gamma'}} \mathbf{S}_{i,\gamma}\cdot\mathbf{S}_{i,\gamma'} +J_{H}\sum_{\substack{i \\ \gamma < \gamma'}}\left( P_{i,\gamma}^{\dagger}P_{\vphantom{i^i}i,\gamma'} + h.c \right), \nonumber \\ \label{Eqn: Multi-Orbital Ham}
\end{eqnarray}

\noindent
where $c^{\dagger}_{i,\gamma,\sigma}$ ($c_{\vphantom{i^i}i,\gamma,\sigma}$) creates (annihilates) an electron at site $i$, with orbital $\gamma$, and spin projection along the $z$-axis $\sigma$. The first term represents the inter- and intra-orbital hopping between only nearest-neighbor sites. General hopping matrices for the two- and three-orbitals per site cases are displayed in Eqs.\eqref{Eqn: Hopping 2-orb} and \eqref{Eqn: Hopping 3-orb}, respectively, and in our study we allowed for the hoppings to vary over broad ranges to search for the largest ratios of Heisenberg interactions. The second term is the standard onsite Hubbard repulsion $U$ between spins $\uparrow$ and $\downarrow$ electrons, at the same orbital. The third term contains the onsite inter-orbital repulsion, with the usual relation $U'=U-2J_{H}$ due to rotational invariance. The fourth term involves the Hund's coupling $J_{H}$ that explicitly shows the ferromagnetic character between orbitals. The last term represents the onsite inter-orbital electron-pair hopping $P_{i,\gamma}=c_{i,\gamma,\uparrow}c_{i,\gamma,\downarrow}$. All these terms in the Hubbard model are canonical. 

The general hopping matrices used here for the exact-diagonalization calculation of two- and three-orbitals per site on the two-site system are: 
\begin{eqnarray}
t^{2-orb}_{\gamma\gamma'} &=& \begin{pmatrix}
t_{11} & t_{12} \\
t_{21} & t_{22}
\end{pmatrix}, \label{Eqn: Hopping 2-orb}\\
t^{3-orb}_{\gamma,\gamma'} &=& \begin{pmatrix}
t_{11} & t_{12} & t_{13} \\
t_{21} & t_{22} & t_{23} \\
t_{31} & t_{32} & t_{33}
\end{pmatrix}, \label{Eqn: Hopping 3-orb}
\end{eqnarray}

\noindent
where $t_{\alpha\beta}$ represents the nearest-neighbor hopping element from orbital $\alpha$ to orbital $\beta$. Due to rotational symmetry of the two-site system, $t_{\alpha\beta}=t_{\beta\alpha}$. This reduces the number of hopping elements from $N_{o}^2$ to $N_{o}(N_{o}+1)/2$, where $N_o$ is the number of orbitals.

\subsection{Heisenberg Model with Higher Order Terms}
The allowed high-order Heisenberg model for any spin-$S$ system can be written generically as:
\begin{equation}
H_S=\sum_{\langle i,j\rangle} \sum_{n=1}^{2S} J_{n}\left(\mathbf{S}_{i}.\mathbf{S}_{j}\right)^{n}.
\end{equation}

Using the above equation we can write the general Hamiltonian for $S=1$ spin system as:
\begin{equation}
H_{1}=\sum_{\langle i,j\rangle} \left[J_{1}\left(\mathbf{S}_{i}.\mathbf{S}_{j}\right) + J_{2}\left(\mathbf{S}_{i}.\mathbf{S}_{j}\right)^{2}\right]. \label{Eqn: S=1 Ham}
\end{equation}

\begin{figure}[!bth]
\centering
\includegraphics[width=3.2in, height=2.7in]{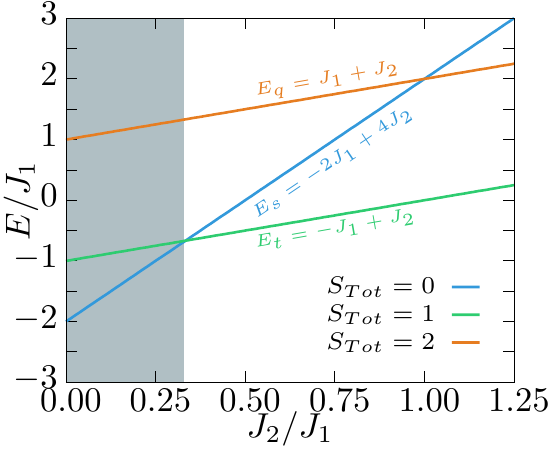}
\caption{Energy ($E/J_1$) vs. $J_2/J_1$ for the two-site $S=1$ Heisenberg model. The shaded area depicts the region with ordering singlet, triplet and quintuplet in increasing order of energies, as it occurs in the more fundamental
two-orbital per site Hubbard model. \label{Fig: E/J1 vs. J2/J1 Plot}}
\end{figure}

We diagonalize the Hamiltonian in Eq.~\eqref{Eqn: S=1 Ham} for the two-site system and obtain the following three energy levels:
\begin{equation}
\begin{rcases*}
E_s = -2J_1+4J_2, & \text{Singlet}~{\it s}\\
E_t = -J_1+J_2, & \text{Triplet}~{\it t} \\
E_q = J_1+J_2, & \text{Quintuplet}~{\it q} \label{Eqn: S=1 Energies}
\end{rcases*}.%
\end{equation}

In Fig.~\ref{Fig: E/J1 vs. J2/J1 Plot}, we illustrate the plot of these energy levels vs. $J_2/J_1$. For $J_2/J_1 < 1/3$, the ordering of these levels strictly follows the singlet-triplet-quintuplet sequence in increasing order of energies. This is vital as the same sequence appears in the more fundamental two-orbital per site Hubbard model in strong coupling.

Of course, when comparing these energies mentioned in Eq.~\eqref{Eqn: S=1 Energies} with the Hubbard results obtained from exact-diagonalization in the strong coupling regime a constant offset in energies must be included, leading generically to $E'_{a}=E_{a}+E_{off}$ where $a=s,t,q$ and $E_{off}$ is the offset energy. Based on this information and the energies provided in Eq.~\eqref{Eqn: S=1 Energies} one can compute the ratio $J_2/J_1$ in terms of the Hubbard energies obtained from exact-diagonalization $E'_{a}$'s as:
\begin{equation}
\frac{J_2}{J_1} = \frac{E'_q -3 E'_t + 2E'_s}{3(E'_q - E'_t)}. \label{Eqn: J2/J1 for 2-Orbs}
\end{equation}

The above equation is used for calculating the values of $J_2/J_1$ in our two-site two-orbitals per site  exact-diagonalization study, in the range where the Hubbard model energies are in the expected singlet-triplet-quintuplet order, starting from the singlet ground state (this assumption tends to break down only in weak coupling, already outside the range of the Heisenberg model description).

Similarly for $S=3/2$ the high-order Heisenberg Hamiltonian reads:
\begin{equation}
H_{\frac{3}{2}}=\sum_{\langle i,j\rangle} \left[J_{1}\left(\mathbf{S}_{i}.\mathbf{S}_{j}\right) + J_{2}\left(\mathbf{S}_{i}.\mathbf{S}_{j}\right)^{2} + J_{3}\left(\mathbf{S}_{i}.\mathbf{S}_{j}\right)^{3}\right]. \label{Eqn: S=3/2 Ham}
\end{equation}

We diagonalize this Hamiltonian in Eq.~\eqref{Eqn: S=3/2 Ham} for the two-site system and obtain four energy levels:
\begin{equation}
\begin{rcases*}
E_s = \frac{-15}{64}\left(16J_1 - 60J_2 + 225J_3 \right), & \text{Singlet}~{\it s}\\
E_t = \frac{-11}{64}\left(16J_1 - 44J_2 + 121J_3 \right), & \text{Triplet}~{\it t}\\
E_q = \frac{-3}{64}\left(16J_1 - 12J_2 + 9J_3 \right), & \text{Quintuplet}~{\it q}\\
E_v = \frac{9}{64}\left(16J_1 + 36J_2 + 81J_3 \right), & \text{Septuplet}~{\it v} \label{Eqn: S=3/2 Energies}
\end{rcases*}.%
\end{equation}

Following the same reasoning as in the case of $S=1$, i.e. considering an offset energy, then $E'_{a}=E_{a}+E_{off}$ where $a=s,t,q,v$, and using the set of equations provided in Eq.~\eqref{Eqn: S=3/2 Energies} the analytical expression for $J_2/J_1$ and $J_3/J_1$ in terms of $E'_{a}$'s for $S=3/2$ becomes
\begin{equation}
\frac{J_2}{J_1} =  \frac{4}{3}\frac{\left(29E'_v - 85E'_q + 81E'_t - 25E'_s\right)}{\left( 81E'_v + 115E'_q - 351E'_{t} + 155E'_{s} \right)}, \label{Eqn: J2/J1 for 3-Orbs}
\end{equation}

\noindent
and
\begin{equation}
\frac{J_3}{J_1} = \frac{16}{3}\frac{\left(E'_v - 5E'_q + 9E'_t - 5E'_s\right)}{\left( 81E'_v + 115E'_q - 351E'_{t} + 155E'_{s} \right)}. \label{Eqn: J3/J1 for 3-Orbs}
\end{equation}

\noindent
Equations~\eqref{Eqn: J2/J1 for 3-Orbs} and~\eqref{Eqn: J3/J1 for 3-Orbs} were used for calculating the values of $J_2/J_1$ and $J_3/J_1$ in our two-site three-orbitals per site exact-diagonalization study, respectively. Here, we do not include a figure like Fig.\ref{Fig: E/J1 vs. J2/J1 Plot} for the case of $S=3/2$ because it would require a three-dimensional plot of energy vs. $J_2/J_1$ and $J_3/J_1$ which would be difficult to visualize. For this reason, we simply have included here the relevant equations that were employed.

\section{Results}
In this section, we will discuss our numerical results via exact-diagonalization for the two-site system. Note that
not only $U$ and $J_H$ will be varied, but the most time-consuming portion of the calculation arises from the large number of hopping
amplitude ratios that were studied (using $t_{11}$ as unit of reference). As a consequence, we have analyzed hundreds of different ratios of
Hamiltonian parameters and in all cases mapped the low-energy results into the corresponding Heisenberg models. 
Specifically, on average we run over
30 values of $U$ and 12 values of $J_H/U$, for a fixed set of hopping amplitudes. This already amounts to 360 runs. 
For two orbitals per site, we used 36 combinations of $t_{22}/t_{11}$ and $t_{12}/t_{11}$ for a total of 360$\times$36 = 12,960 cases. 
For three orbitals per site, we used 196 combinations of the ratios
$t_{22}/t_{11}$, $t_{33}/t_{11}$, $t_{12}/t_{11}$, $t_{13}/t_{11}$, and $t_{23}/t_{11}$ for a total of 360$\times$196=70,560 cases.
Crudely estimating, the total number of cases studied is of the order of 4$\times$10$^{4}$. We computationally automatized the fittings, and from the vast array of numbers we isolated
approximately 150 sets of data containing the largest ratios for $J_2/J_1$ and $J_3/J_1$. Those special cases were plotted and visually inspected. From that set, the very small subset displayed in the figures shown in the present publication is the subset that in our
judgement best represents the cases where the Heisenberg coupling ratios are the largest in absolute value, because our primary aim is to establish upper bounds on those quantities. These ratios can be positive or negative.


\subsection{Two-Site Two-Orbitals per site \label{Sec: 2-site 2-ors}}

\begin{figure}[H]
\centering
\includegraphics[width=3.2in, height=2.7in]{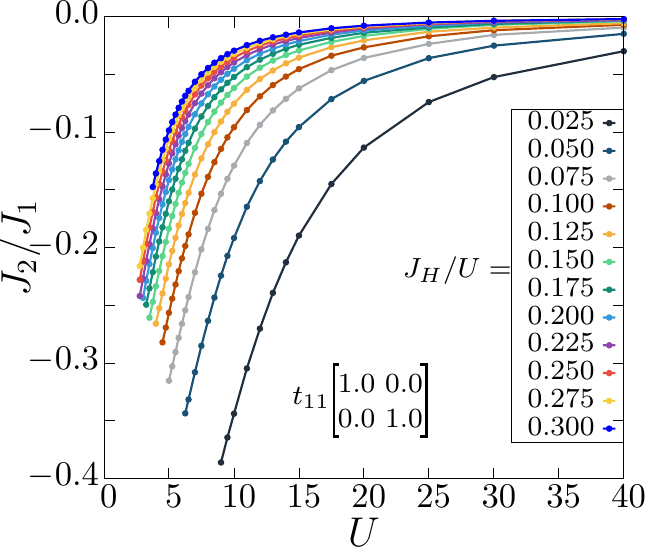}
\caption{$J_2/J_1$ plotted vs. $U$ for a two-site two-orbitals per site system via exact-diagonalization, at the various ratios $J_H/U$ indicated. The hopping parameters chosen are $t_{22}=t_{11}$ and $t_{12}=t_{21}=0$ for this example, namely the unit matrix. 
To help the readers, the hopping matrix is presented as inset in the plot. The bandwidth 
for this set of hopping parameters is $W=4t_{11}$. Note that ``bandwidth'' is defined with regards
to the tight-binding model with the hoppings used here but in the bulk limit.   \label{Fig: J2/J1 vs. U Plot t22=t11, t12=0}}
\end{figure}

Here, we present our two-site two-orbitals per site exact-diagonalization results. All the results presented below have the same low-energy order: first a singlet (with the total spin $S_{Tot}=0$) for the ground state, then a triplet ($S_{Tot}=1$) for the first excited state, and finally a quintuplet ($S_{Tot}=2$) for the second excited state.

\begin{figure}[!t]
\centering
\includegraphics[width=3.2in, height=2.7in]{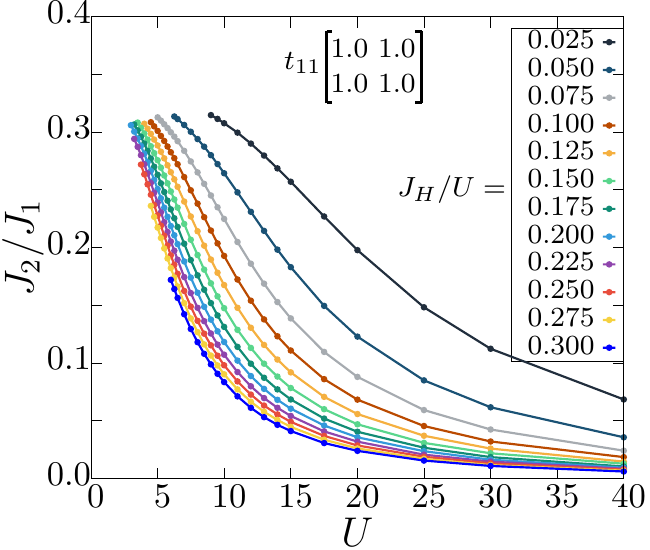}
\caption{$J_2/J_1$ plotted vs. $U$ for a two-site two-orbital per site system via exact diagonalization, at the various ratios $J_H/U$ indicated. The hopping parameters chosen are $t_{22}=t_{12}=t_{21}=t_{11}$. This hopping matrix is presented in the plot as inset. The bandwidth for these set of hopping parameters is $W=8t_{11}$
(for definition of bandwidth see caption of Fig.~\ref{Fig: J2/J1 vs. U Plot t22=t11, t12=0}). 
\label{Fig: J2/J1 vs. U Plot t22=t11=t12}}
\end{figure}

In both Fig.~\ref{Fig: J2/J1 vs. U Plot t22=t11, t12=0} and Fig.~\ref{Fig: J2/J1 vs. U Plot t22=t11=t12} we first performed exact diagonalization of the multi-orbital Hubbard model defined in Eq.~\eqref{Eqn: Multi-Orbital Ham}. The hopping parameters used for each of these figures is shown as an inset, for better visualization, 
and also in the caption. As already explained, for each $J_H/U$, we identified the range of $U$  that gives us the ordering: singlet, triplet and quintuplet for the ground-state, first excited-state, and second excited state, respectively. After the proper regime of couplings was identified, the energies of these respective states were used to calculate the ratio $J_2/J_1$ using Eq.\eqref{Eqn: J2/J1 for 2-Orbs}. 

Our main result is that the largest ratio observed (in absolute value) is close to 0.4. For a wide variety of ``less symmetric'' 
hopping amplitudes, namely employing neither the unit matrix or the matrix with all elements equal, we
observed that $|J_2/J_1|$ is smaller than those reported in Figs.~\ref{Fig: J2/J1 vs. U Plot t22=t11, t12=0} and ~\ref{Fig: J2/J1 vs. U Plot t22=t11=t12}. 
Two important details to remark are: (a) the ratios shown can be both positive and negative and for this reason the two examples shown were chosen. In both cases,
positive and negative, the largest magnitudes of the ratios are not too different. (b) As obvious from the figures, the largest ratios are obtained as $U$ is reduced from very strong coupling. This makes sense because in the limit where a perturbative expansion in $t_{11}/U$ is valid, $J_1$ is the lowest order and $J_2$ the next leading order. Naturally, their ratio of coefficients scales as $t_{11}/U$ and $J_2/J_1$ converges to zero as $U$ diverges. As a consequence, we can firmly conclude that the most promising region to observe the effects of the biquadratic term is the range of $U/W \sim 1$, which is the {\it intermediate coupling} regime. This region of parameter space usually contains a variety of exotic phases because here several tendencies are in close competition leading
to ``frustration'' effects which are hidden, namely not obvious at the Hamiltonian level.

\onecolumngrid

\begin{figure}[H]
\centering
\includegraphics[width=6in, height=3in]{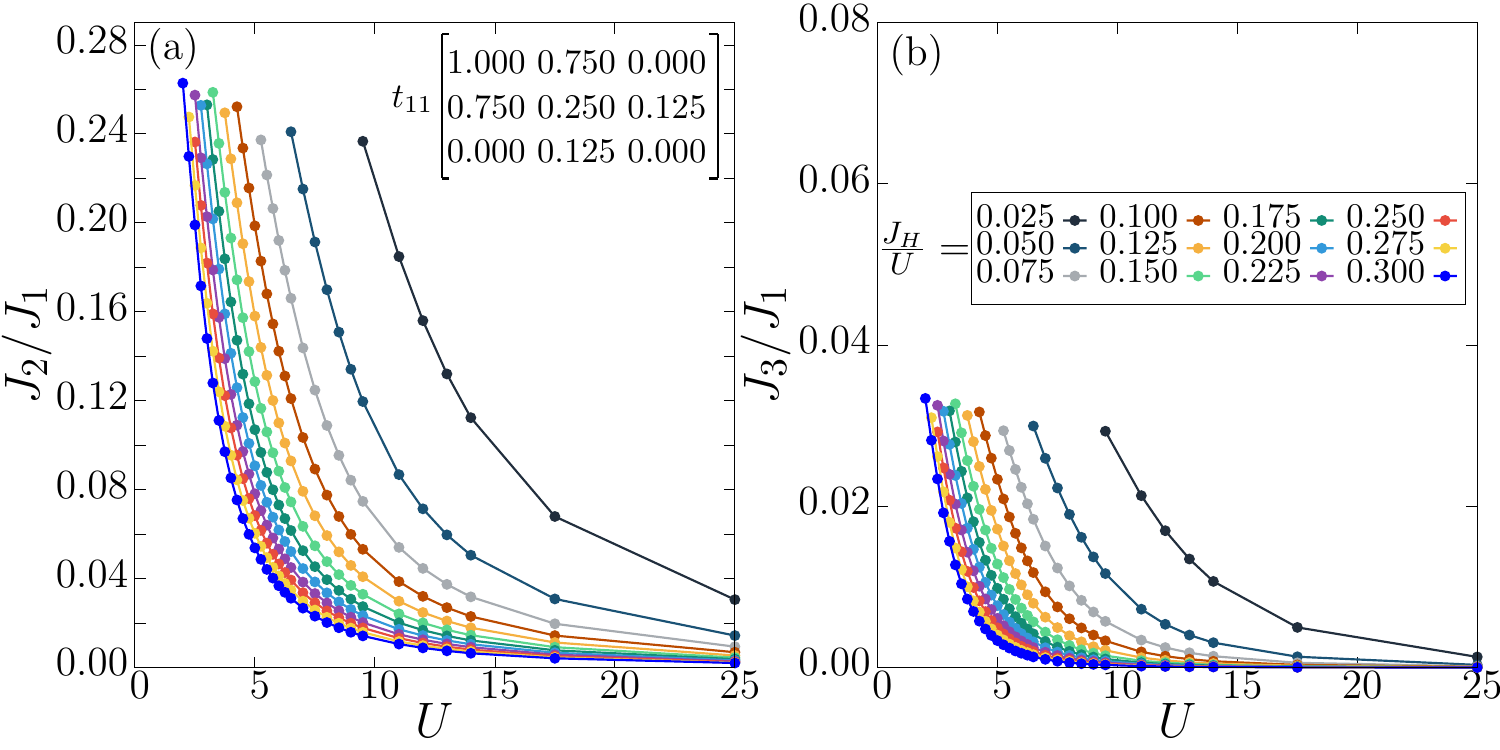}
\vspace*{-0.5cm}
\caption{(a) $J_2/J_1$ and (b) $J_3/J_1$ vs. $U$ for the two-site three-orbitals per site system obtained via exact diagonalization, at the several values of $J_H/U$ indicated. The hopping parameters chosen for this particular plot are $t_{11}=1.0$, $t_{22}=0.25$, $t_{33}=0$, $t_{12}=0.75$, $t_{23}=0.125$ and $t_{13}=0$. They are also 
shown in an inset in the figure for the benefit of the readers. The bandwidth for this particular set of hopping parameters is $W\approx 5.86t_{11}$ (for definition of bandwidth see caption of Fig.~\ref{Fig: J2/J1 vs. U Plot t22=t11, t12=0}). The color convention for the ratio $J_H/U$, as well as the hopping matrix, is common to both panels. \label{Fig: J2/J1 and J3/J1 vs. U Plot t22=0.25t11, t33=t13=0, t12=0.75t11, t23=0.125t11}}
\end{figure}

\twocolumngrid

\subsection{Two-Site Three-Orbitals per site}

In this subsection, we present our two-site three-orbitals per site exact diagonalization results. All the results presented below have the same energy ordering, namely singlet ($S_{Tot}=0$) for the ground state, triplet ($S_{Tot}=1$) for the first-excited state, quintuplet ($S_{Tot}=2$) for the second-excited state, and septuplet ($S_{Tot}=3$) for the third-excited state. The latter originates in the three orbital per site nature of the problem, and it does not appear for two orbitals per site. Namely, the extra spin manifold occurs because the total number of electrons in the system is 6 which allows total spins 3, 2, 1, and 0, contrary to a total of 4 electrons in the previous subsection.

Unlike the two-site two-orbital per site case, here we observe that it is the ``less symmetric" (as mentioned in section \ref{Sec: 2-site 2-ors}) hopping amplitudes that gives large values of the ratios $|J_2/J_1|$ and $|J_3/J_1|$.

\onecolumngrid

\begin{figure}[H]
\centering
\includegraphics[width=6in, height=3in]{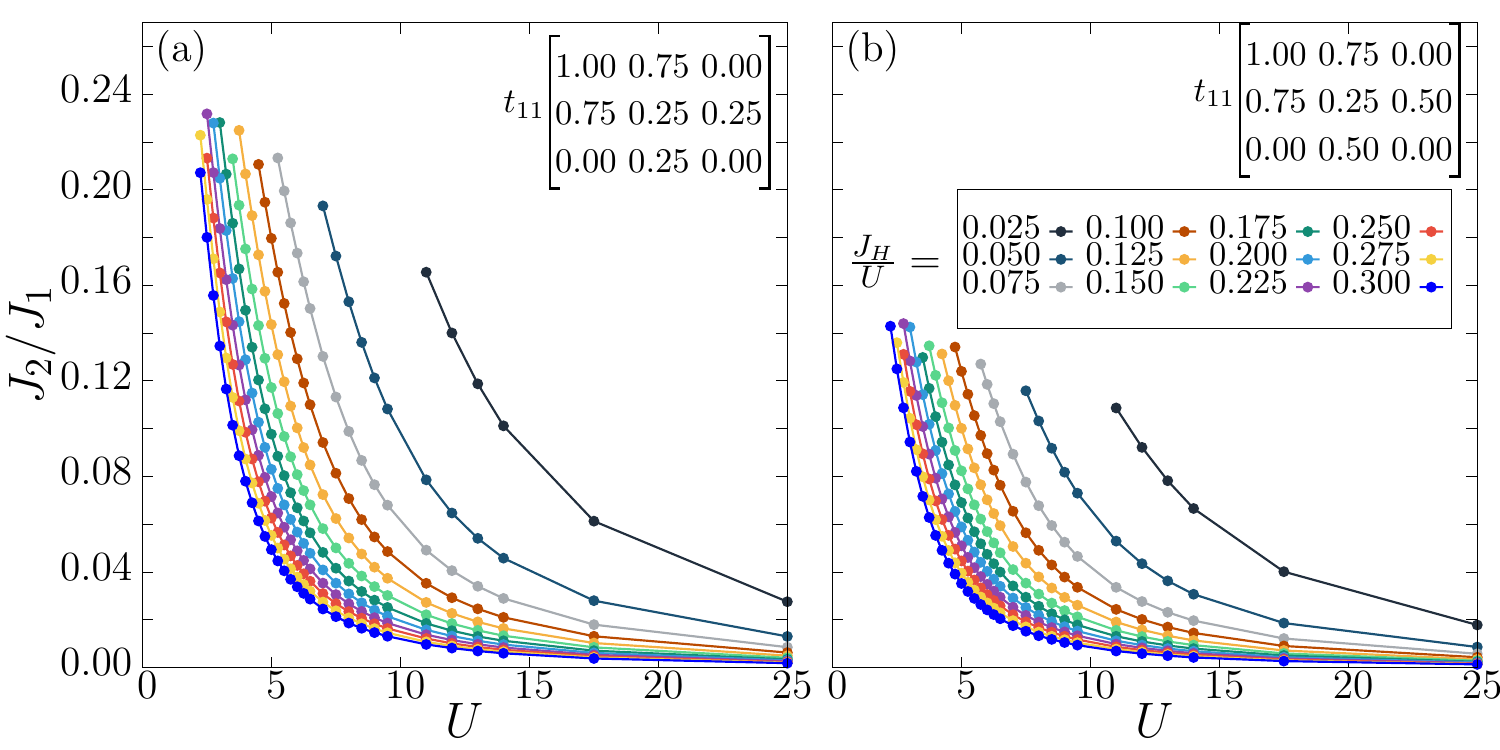}
\vspace*{-0.5cm}
\caption{$J_2/J_1$ vs. $U$ for a two-site three-orbitals per site system obtained 
via exact diagonalization, at the several 
ratios $J_H/U$ shown. The color convention is the same in (a) and (b). The hopping parameters chosen for panel (a) are $t_{11}=1.0$, $t_{22}=0.25$, $t_{33}=0$, $t_{12}=0.75$, $t_{23}=0.25$ and $t_{13}=0$, with bandwidth $W\approx5.9t_{11}$ and in (b) are $t_{11}=1.0$, $t_{22}=0.25$, $t_{33}=0$, $t_{12}=0.75$, $t_{23}=0.5$ and $t_{13}=0$, with bandwidth $W\approx 6.05t_{11}$ (for definition of bandwidth see caption of Fig.~\ref{Fig: J2/J1 vs. U Plot t22=t11, t12=0}). The hopping matrices are shown also in each panel for the benefit of the readers. The
ratios $J_3/J_1$ are not shown because they are considerably smaller than $J_2/J_1$, as in Fig.~\ref{Fig: J2/J1 and J3/J1 vs. U Plot t22=0.25t11, t33=t13=0, t12=0.75t11, t23=0.125t11} . \label{Fig: J2/J1 vs. U Plot t22=0.25t11, t33=t13=0, t12=0.75t11, t23=0.25t11 & 0.5t11}}
\end{figure}

\twocolumngrid

Qualitatively, the conclusions in Fig.~\ref{Fig: J2/J1 and J3/J1 vs. U Plot t22=0.25t11, t33=t13=0, t12=0.75t11, t23=0.125t11} 
resemble those for the two-orbital per site case. Once again, the ratios are the largest as $U/t_{11}$ decreases from the strong coupling regime. Thus, again
we conclude that the intermediate coupling region $U/W \sim 1$ is the most promising to observe sizable values for $J_2$ and $J_3$. Also, the largest values of $J_2/J_1$ are similar to those of the two-orbital per site case. However, as expected from the strong coupling expansion, 
$J_3/J_1$ is an order of magnitude smaller than $J_2/J_1$ because it requires the next order in the large $U$ expansion to develop as compared with $J_2/J_1$.

Figure~\ref{Fig: J2/J1 vs. U Plot t22=0.25t11, t33=t13=0, t12=0.75t11, t23=0.25t11 & 0.5t11} illustrates the dependence of the results varying slightly the hopping amplitudes. Focusing on the matrices contained in both panels, the only difference between the two cases resides in 
the hopping $t_{23}$, which varies by a factor 2. However, this relatively small modification leads to a reduction in approximately a factor two in the
values of $J_2/J_1$. This high sensitivity to small changes in the hoppings is surprising. Such effect manifest the most at intermediate couplings, while
in strong coupling the ratios are less sensitive to small hopping modifications. 

In Figure~\ref{Fig: J2/J1 and J3/J1 vs. U Plot t22=t11, t33=tab=0}, we illustrate the case where the hoppings reside only along the diagonal, but one of them, i.e. $t_{33}$, is zero.
Surprisingly, in this case the fits lead to negative values for both $J_2/J_1$ and $J_3/J_1$. The strength is also reduced when compared with 
Fig.~\ref{Fig: J2/J1 vs. U Plot t22=0.25t11, t33=t13=0, t12=0.75t11, t23=0.25t11 & 0.5t11}. In Fig.~\ref{Fig: J2/J1 vs. U Plot t22=t11, t33=0.125t11 & 0.25t11, tab=0},
$J_2/J_1$ is shown now increasing $t_{33}$ from zero, as compared with Fig.~\ref{Fig: J2/J1 and J3/J1 vs. U Plot t22=t11, t33=tab=0}. Here we see that as we increase $t_{33}$ the  largest value of the ratio $J_2/J_1$ decreases slowly, indicating that in order to find the maximum possible value of the ratio $J_2/J_1$ the hopping amplitude $t_{33}$ must be zero. Similarly, we tune other hopping amplitudes and find the best possible scenario where we achieve the largest value of $J_2/J_1$ and $J_3/J_1$. 

\onecolumngrid

\begin{figure}[H]
\centering
\includegraphics[width=6in, height=2.8in]{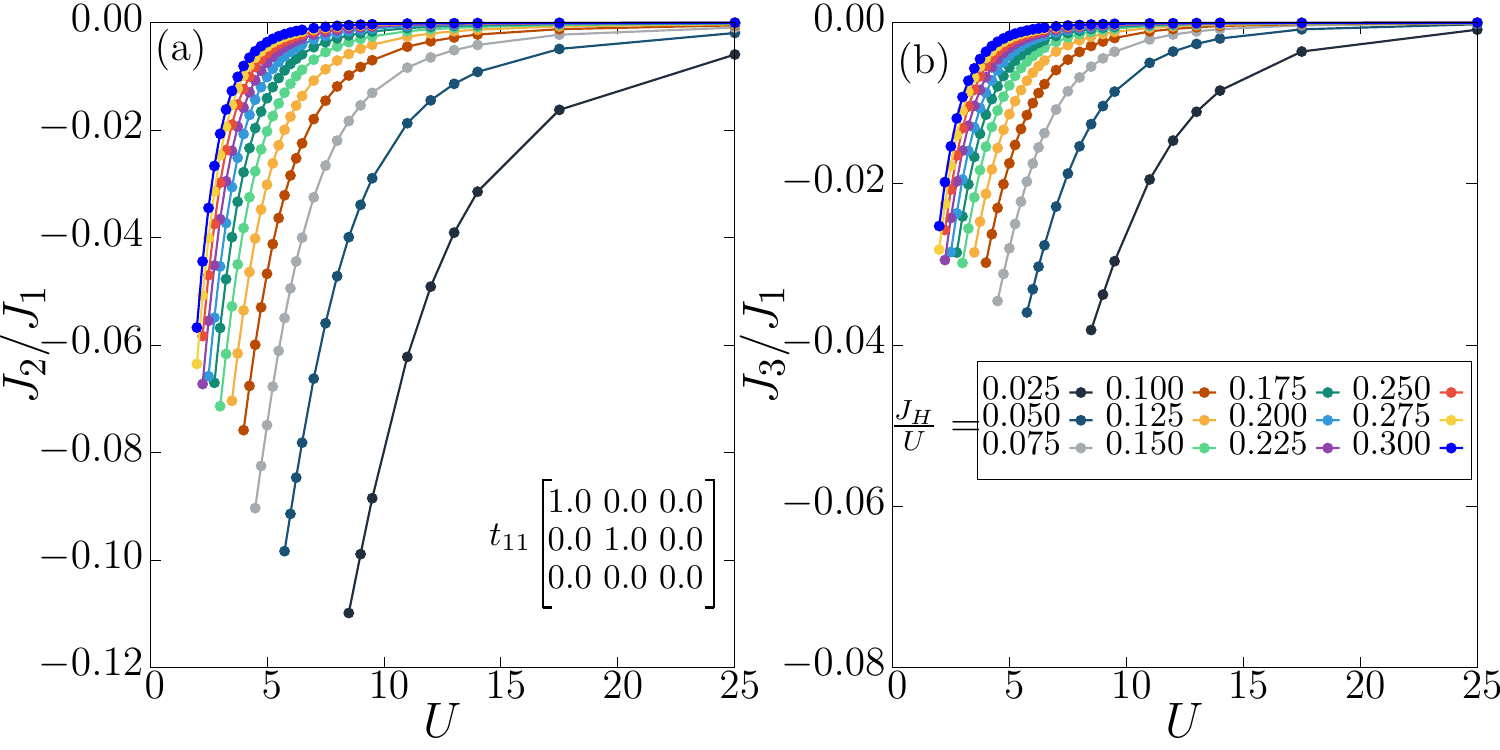}
\vspace*{-0.5cm}
\caption{(a) $J_2/J_1$ and (b) $J_3/J_1$ plotted vs. $U$ for a two-site three-orbitals per site system obtained via exact diagonalization, at the several values of $J_H/U$ indicated. Color convention is common to both panels. The hopping parameters chosen for the plot are $t_{11}=t_{22}=1.0$, $t_{33}=0$ and $t_{\alpha\beta}=0$ for all $\alpha\neq\beta$. The bandwidth for these set of hopping parameters is $W=4t_{11}$ (for definition of bandwidth see caption of Fig.~\ref{Fig: J2/J1 vs. U Plot t22=t11, t12=0}).  \label{Fig: J2/J1 and J3/J1 vs. U Plot t22=t11, t33=tab=0}}
\end{figure}

\twocolumngrid

\onecolumngrid

\begin{figure}[H]
\centering
\includegraphics[width=5.9in, height=2.9in]{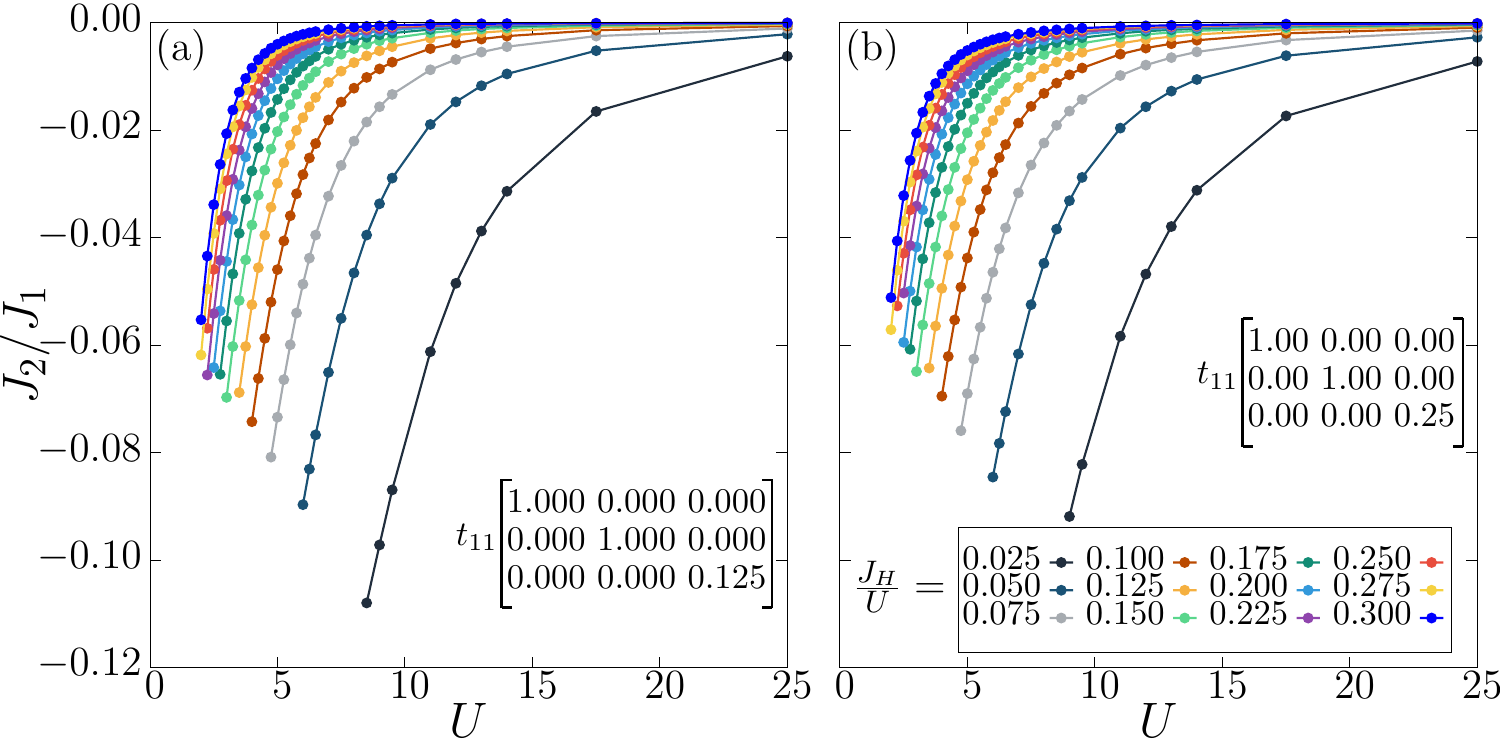}
\vspace*{-0.5cm}
\caption{$J_2/J_1$ plotted vs. $U$ for a two-site three-orbitals per site system using exact diagonalization, at the indicated several ratios of $J_H/U$ (color convention is the same for both panels). The hopping parameters chosen for panel (a) are $t_{11}=t_{22}=1.0$, $t_{33}=0.125$ and $t_{\alpha\beta}=0$ for all $\alpha\neq\beta$ and for panel (b) are $t_{11}=t_{22}=1.0$, $t_{33}=0.25$ and $t_{\alpha\beta}=0$ for all $\alpha\neq\beta$. These hoppings are also shown in the insets. The bandwidth for both sets of hopping parameters is $W=4t_{11}$ (for definition of bandwidth see caption of Fig.~\ref{Fig: J2/J1 vs. U Plot t22=t11, t12=0}). 
\label{Fig: J2/J1 vs. U Plot t22=t11, t33=0.125t11 & 0.25t11, tab=0}}
\end{figure}

\twocolumngrid

\section{Discussion}

In our study we have focused on a two-site electronic multi-orbital Hubbard model to deduce what 
range of Heisenberg couplings are possible at intermediate and large values of the interaction strength $U$. In particular, for two orbitals per site
we focused on how large the biquadratic coupling strength $J_2$ can become, in terms of the canonical quadratic Heisenberg 
superexchange interaction $J_1$. We found that $J_2/J_1$ can be of both signs, but in magnitude 
cannot exceed $\sim$0.4. This allows for the model used by Affleck and collaborators~\cite{affleck} to be realized
employing electronic models. It would be interesting to investigate if this electronic model -- namely selecting 
suitable Hubbard $U$, Hund coupling $J_H$, and hoppings such that $J_2/J_1=1/3$ -- will also lead to a valence bond ground state, although likely the said electronic model will not be exactly solvable. On the other hand, the
exactly solvable case $J_2/J_1=1$ cannot be realized with the model we used.
For the case of spin $S=3/2$, namely using three orbitals per site, the conclusions are similar: once again $J_2/J_1$
cannot exceed $\sim$0.4, while $J_3/J_1$ is even smaller by a factor approximately two.

Our study also suggests that some spin-only models that were studied to search for quantum spin liquids 
must impose constraints on the parameter space explored. To realize spin liquids using electronic models
the most optimal path continues being the addition of 
hoppings beyond nearest-neighbors to create explicit frustration.

Note that our conclusions are in agreement with calculations using 
a two-orbital per site Hubbard model~\cite{tanaka2018}, carried out perturbatively at small $t/U$ 
up to fourth order (first and third order cancel; the second order gives the canonical quadratic Heisenberg model, 
and the fourth order is the one relevant for our discussion involving the biquadratic contribution). 
By this fairly different procedure, nevertheless a conclusion similar to ours was reached: the ratio $J_2/J_1$ is severely limited at large $U$.
Our results, valid at any value of $U$ because they do not rely on perturbation theory, 
suggest that intermediate $U$ is more promising than strong $U$, but still $J_2/J_1$ cannot reach values above 0.4.
The methodology proposed in Ref.~\cite{mila2000}, adding to the problem 
an extra orbital residing in a neighboring site,
may reduce $J_1$, providing a promising path to enhance the ratio $J_2/J_1$~\cite{lin2021}. Our next goal is to investigate the effect of spin-orbit coupling in these two-site multi-orbital Hubbard systems~\cite{nitin2020} and estimate the range of biquadratic-quadratic Heisenberg couplings.

\subsection*{Acknowledgments} 
R.S., N.K., F.R., A.M., and E.D. were supported by the U.S. Department of Energy (DOE), Office of Science, Basic Energy Sciences (BES), Materials Sciences and Engineering Division.

\FloatBarrier

\end{document}